\documentclass[aip,apl,reprint,groupedaddress,superscriptaddress]{revtex4-1}
\usepackage[english]{babel}
\usepackage[utf8x]{inputenc}
\usepackage{amssymb}
\usepackage{amsthm}
\usepackage{amsmath}
\usepackage{hyperref}
\usepackage{subfigure}
\usepackage[percent]{overpic} 
\usepackage{picture}
\usepackage{color}
\usepackage{nth}
\usepackage{gensymb}
\usepackage{cleveref}
\usepackage{textcomp}
\usepackage{makecell}
\usepackage{lineno} 
\usepackage{textcomp}
\usepackage{xcolor}
\graphicspath{{img/}}

\begin{document}

\title{Guiding of charged particle beams in curved capillary-discharge waveguides}

\author{R. Pompili}
\email[]{riccardo.pompili@lnf.infn.it}
\affiliation{Laboratori Nazionali di Frascati, Via Enrico Fermi 40, 00044 Frascati, Italy}
\author{G. Castorina}
\affiliation{University of Rome Sapienza, Piazzale Aldo Moro 5, 00185 Rome, Italy}
\affiliation{University of Catania, Piazza Università 2, 95131 Catania, Italy}
\author{M. Ferrario}
\author{A. Marocchino}
\affiliation{Laboratori Nazionali di Frascati, Via Enrico Fermi 40, 00044 Frascati, Italy}
\author{A. Zigler}
\affiliation{Racah Institute of Physics, Hebrew University, 91904 Jerusalem, Israel}

\date{\today}

\begin{abstract}
A new method able to transport charged particle beams along curved paths is presented. It is based on curved capillary-discharge waveguides in which the induced azimuthal magnetic field is used both to focus the beam and keep it close to the capillary axis.
We show that such a solution is highly tunable, it allows to develop compact structures providing large deflecting angles and, unlike conventional solutions based on bending magnets, preserves the beam longitudinal phase space. The latter feature, in particular, is very promising when dealing with ultra-short bunches for which non-trivial manipulations on the longitudinal phase spaces are usually required when employing conventional devices. 
\end{abstract}


\keywords{}

\maketitle


Nowadays there is a growing interest in the development of new devices able to deflect charged particle beams with ever greater energies.
In accelerator facilities, dipole magnets are widely used to realize bends and translate the beam to a specific location~\cite{reiser,bryant2005principles,leemann2009beam} and for the generation of synchrotron radiation~\cite{sokolov1966synchrotron,helliwell1998synchrotron}, with a broad range of applications~\cite{brown2002overview,weik2000specific,lewis1997medical,suortti2003medical}. 
So far permanent and electromagnetic devices have been extensively employed although different solutions, e.g. based on channeling~\cite{gemmell1974channeling,dabagov2015channeling}, have been proposed.
Superconducting magnets~\cite{tomita2003high} are at the edge of present technology and NbTi superconductors cooled at cryogenic temperatures represent the state of the art. They are routinely used, for instance, at the Large Hadron Collider (LHC) where more than 2/3 of the 27~km-long ring is filled by 8~T superconducting dipole magnets~\cite{todesco2004steering,evans2008lhc} in order to bend the 7~TeV proton beams. 
With current limits of superconducting technology, it is difficult to envision compact solutions that can be scaled to even greater energies and/or larger deflection angles.

\begin{figure}[h]
\centering
\includegraphics[width=85mm]{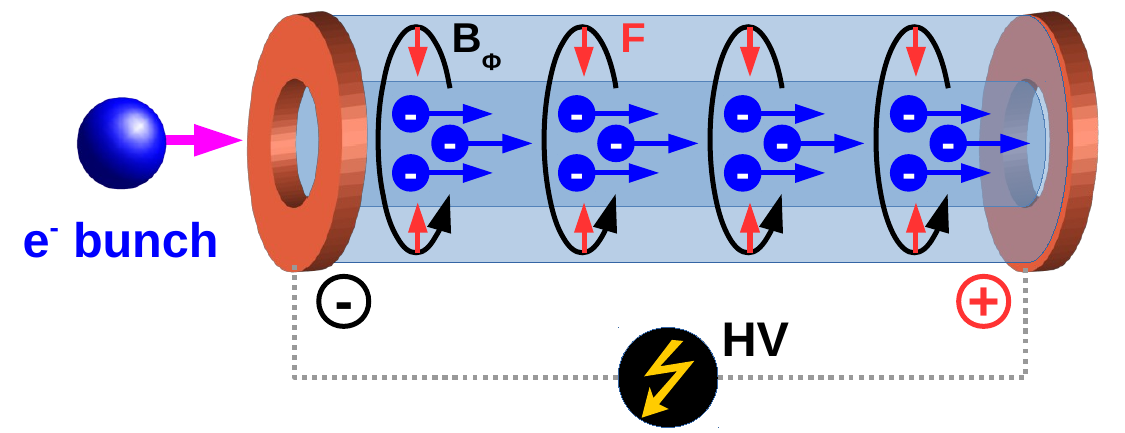}
\caption{Scheme for particle guiding. A discharge current flows through the plasma capillary inducing, according to the Ampere law, an azimuthal magnetic field $\mathbf{B}_{\phi}$ whose resulting force \textbf{F} focuses an externally injected electron bunch.}
\label{PlasmaLens_sketch}
\end{figure}

In this context plasmas represent an alternative approach. They can sustain extremely high fields~\cite{1979PhRvL..43..267T} and currents~\cite{rocca1993fast,rocca1996energy,tomasel1997lasing,hosokai2000optical} and, for these reasons, today they are replacing conventional accelerators~\cite{leemans2006gev,2007Natur.445..741B} and standard focusing devices~\cite{su1990plasma,boggasch1992plasma,hairapetian1994experimental}.
Recently, several proof of principle experiments have been performed in focusing electron beams by means of the so-called active plasma lenses~\cite{van2015active,pompili2017experimental}, schematically depicted in Fig.~\ref{PlasmaLens_sketch}. They consist of gas-filled capillaries in which the plasma is produced by an electrical discharge~\cite{butler2002guiding,spence2003gas}. According to the Ampere law, an azimuthal magnetic field $B_{\phi}$ is induced whose strength increases radially and, unlike standard quadrupoles, provides symmetric focusing of the beam in both transverse planes.
Here we show that the same mechanism can also be used to deflect particles. In the active plasma lens the Lorentz force produced by the azimuthal magnetic field pushes the travelling particles toward the capillary axis (where it vanishes) and the same applies for curved shapes since the the flux of plasma electrons driven by the discharge actually follows the capillary geometry~\cite{ehrlich1996guiding,reitsma2008propagation}.
Such a device, hereinafter called Active Bending Plasma (ABP), presents interesting features when considering the effects of deflection on the particle beam and can in principle reach higher magnetic fields (and, thus, larger deflection angles) than superconducting magnets.
A detailed study is presented and discussed by means of numerical simulations both for the particle guiding and plasma dynamics.

\begin{figure*}[!t]
\centering
\subfigure{
\begin{overpic}[height=0.37\textwidth]{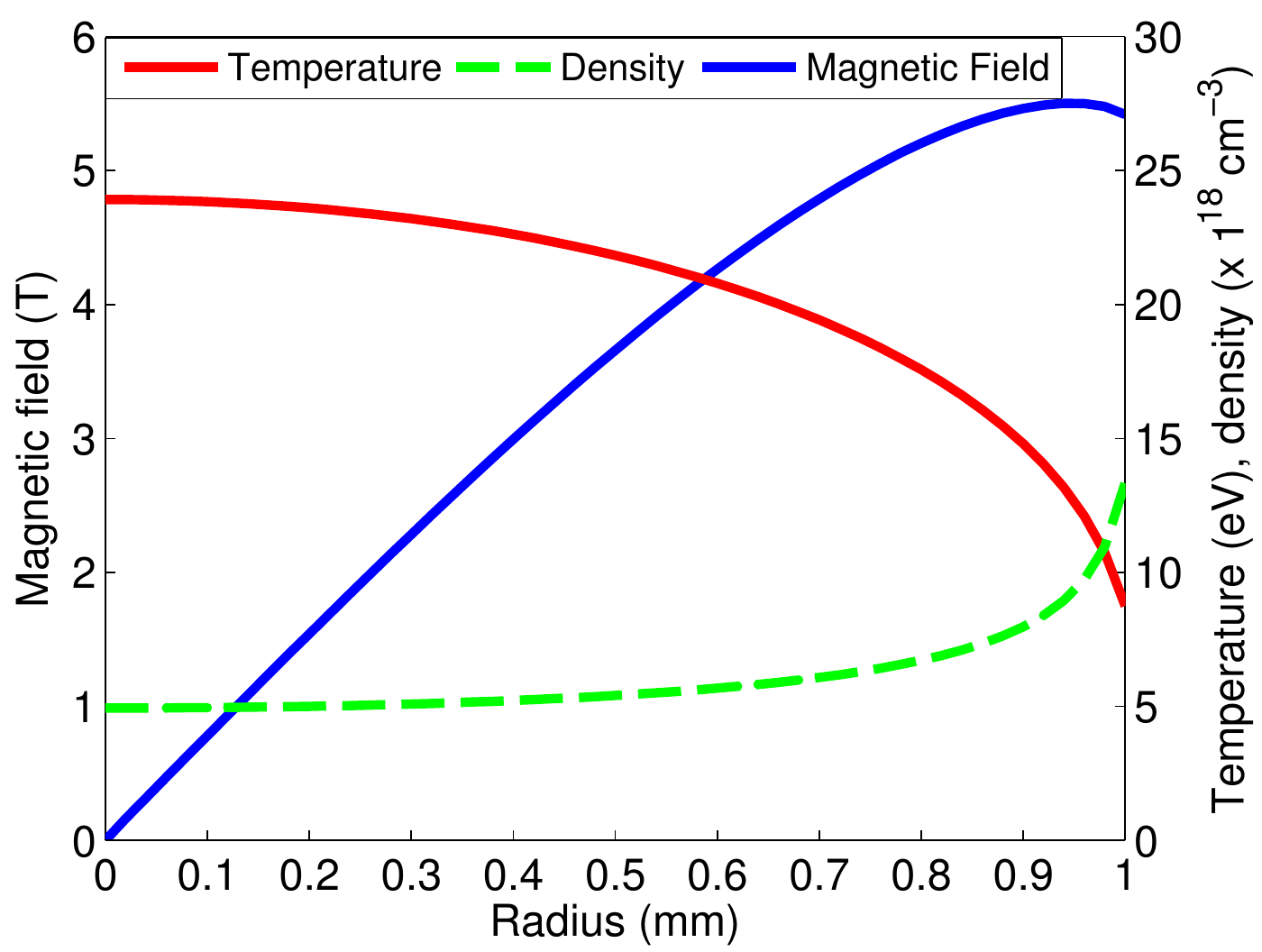}
\put(12,63){\color{black}\textbf{a}}
\end{overpic}
\label{calcProfiles}
}
\subfigure{
\begin{overpic}[height=0.37\textwidth]{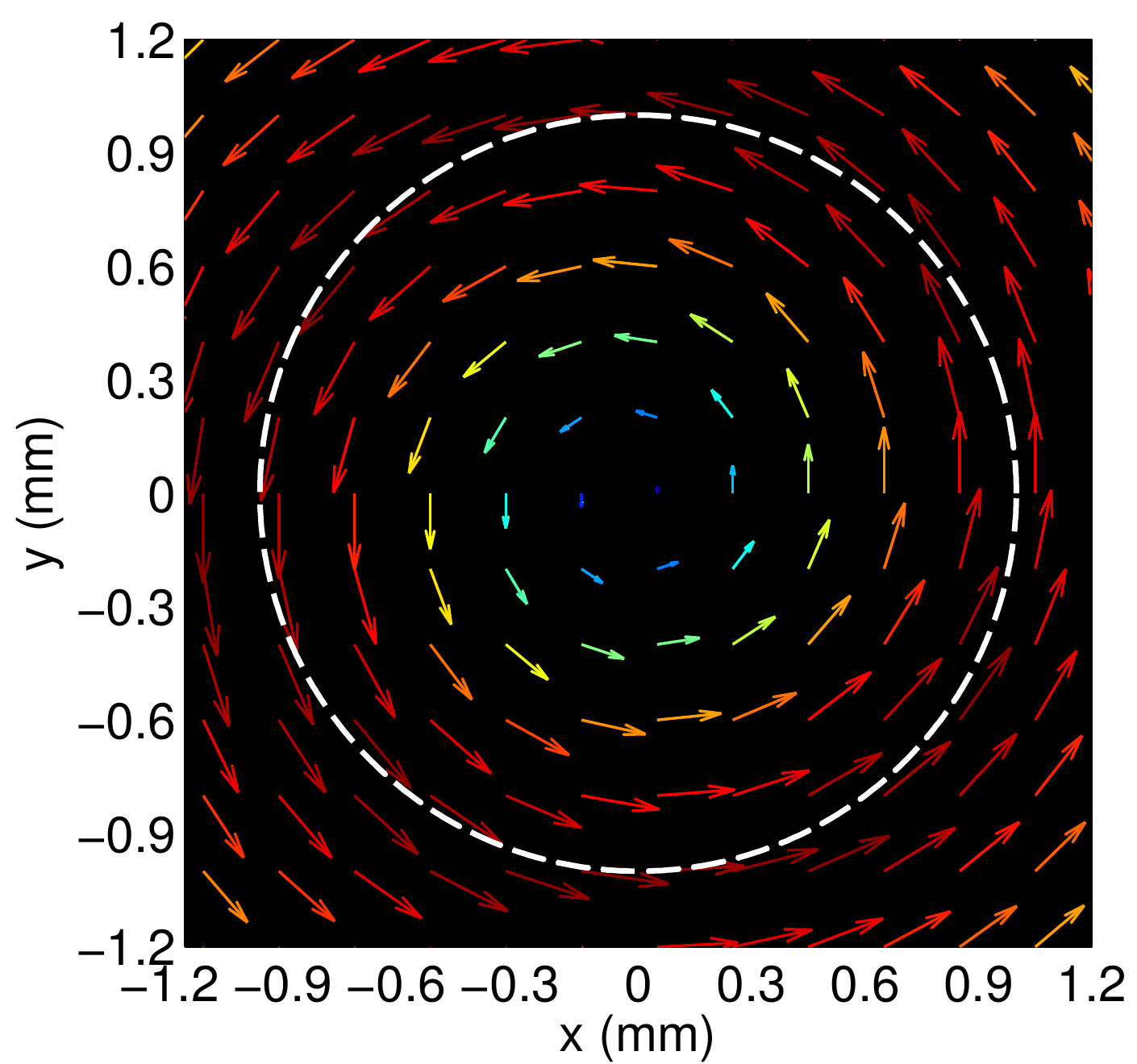}
\put(18,84){\color{white}\textbf{b}}
\end{overpic}
\label{quiverLinear10_cm}
}
\subfigure{
\begin{overpic}[height=0.425\textwidth]{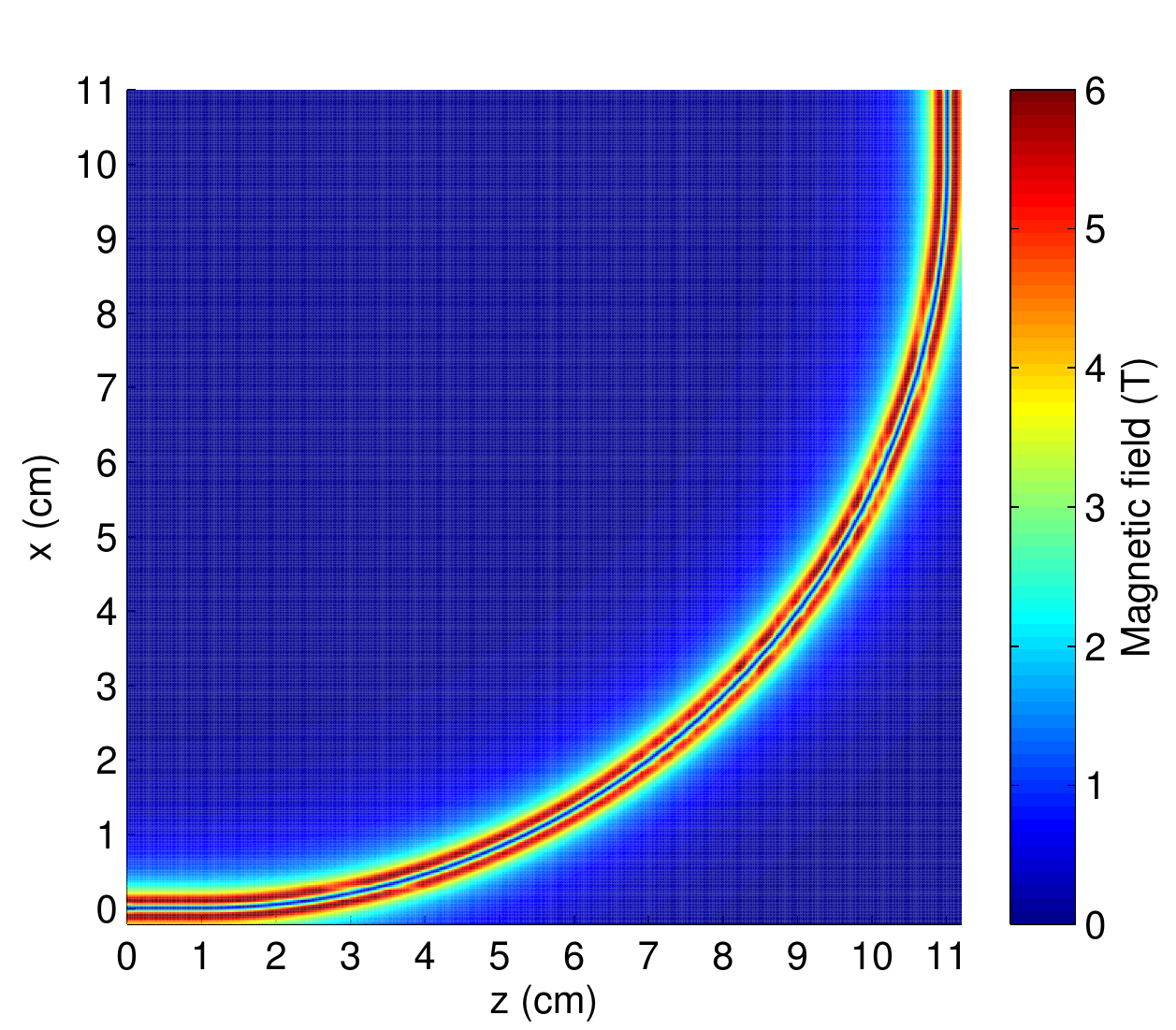}
\put(14,74){\color{white}\textbf{c}}
\end{overpic}
\label{FieldMapLinear10_cm}
}
\subfigure{
\begin{overpic}[height=0.40\textwidth]{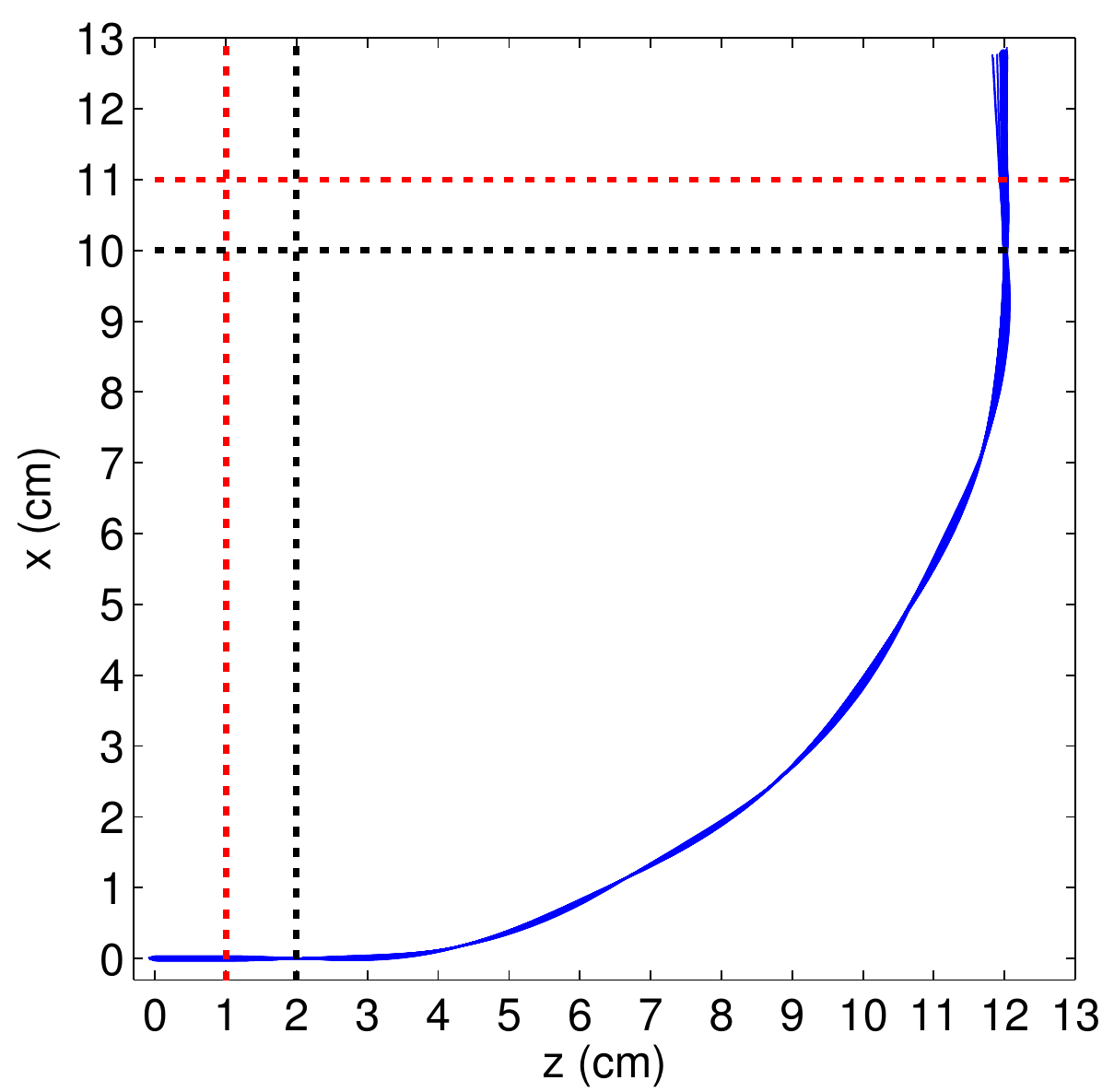}
\put(15,88){\color{black}\textbf{d}}
\end{overpic}
\label{traj}
}
\caption{(a) Calculated radial profiles of the magnetic field (blue), temperature (red) and electron density (green) in a prefilled Hydrogen capillary for $I_D=25$~kA discharge current. (b) Vector plot showing the field lines in the transverse \textit{x-y} plane. The white dashed circle indicates the capillary walls. (c) Magnetic field intensity (along the $y=0$ plane) for the bent capillary with $\rho=10$~cm curvature radius. Two 1~cm-long straight sections are included before and after the bent path. (d) Trajectories of the travelling electrons along the curved path. The red (black) dotted lines show the capillary channel (without) including the two straight sections. A drift of 1~(2)~cm is considered in the simulation upstream (downstream) the capillary channel.}
\label{maps}
\end{figure*}

To test the guiding efficiency of the ABP, in the following we will refer to a reference electron beam with 50~pC charge, energy $E=100$~MeV ($0.1\%$ energy spread), 1~ps duration (or, equivalently, $\sigma_z\approx 300~\mu m$ length), $\epsilon_n=1~\mu m$ normalized emittance and $\sigma_{x,y}=100~\mu m$ spot size\footnote{All the quantities are quoted as root mean square (rms).}.
The particle trajectories are reconstructed by using the General Particle Tracer (GPT) code~\cite{de1996general} while the bunch dynamics along the plasma channel is investigated with Architect, a hybrid-kinetic fluid code~\cite{marocchino2016efficient}.
Similarly to the active plasma lens, the ABP requires a magnetic focusing field whose strength increases with radius. As soon as the particles move away from the axis along the curved path, they experience a larger and larger field pushing them back toward the center of the capillary.
Such a magnetic field can be obtained with so-called capillary discharge waveguides~\cite{butler2002guiding,spence2003gas} in which a discharge current is pulsed through a capillary prefilled with gas. 
Under these assumption, we tested the feasibility of the ABP concept by employing a bent capillary with $R_c=1$~mm hole radius and $\rho=10$~cm radius of curvature. 
With such parameters, for the bending it is required that the magnetic field is at least $B_{\phi}= \beta E c/q_e \rho\approx 3.3$~T, where $\beta$ is the bunch velocity normalized to the speed of light $c$ and $q_e$ is the electron charge. 
The magnetic field radial profile $B_{\phi}(r)$ across the capillary is calculated with a one-dimensional analytical model that assumes the plasma distribution as a balance between Ohmic heating and cooling due to electron heat conduction~\cite{bobrova2001simulations}.
The model computes the plasma temperature profile $T(r)$ allowing to retrieve its electric conductivity $\sigma_e (r)$ and, in turn, the current density $J(r)$ flowing through it. The resulting magnetic field is then calculated according to the Ampere law as $B_{\phi}(r) = (\mu_0 /r) \int_0^r J(r') r' dr'$, with $\mu_0$ the vacuum permeability.

Fig.~\ref{calcProfiles} shows the radial magnetic field that results by applying a discharge current of $I_D=25$~kA to a capillary prefilled by pure Hydrogen gas with $10^{19}$~cm$^{-3}$ density. On the same plot there are also reported the temperature and density profile across the capillary.
The resulting vector plot of the magnetic field lines is shown in Fig.~\ref{quiverLinear10_cm}.
Both the gas and its density have been chosen in order to avoid plasma pinching and maintain a pure Ohmic regime. In these conditions the average plasma temperature and electron density are indeed $T_e\approx 17$~eV and $n_e\approx 7\times 10^{18}$~cm$^{-3}$, respectively. Being the plasma thermal pressure (per volume unit) $p_T= 4 N_e k_B T_e$ (with $k_B$ is the Boltzmann constant and $N_e = n_e \pi R_c^2$ the number of electrons per unit length) and the magnetic pressure $p_B= \mu_0 I_D^2/4\pi$ it is $p_T\gg p_B$ and thus no pinching is expected~\cite{bennett1955self}.
For the sake of completeness we have numerically cross-checked such a statement by employing the so-called "snow-plow" model~\cite{vrba2000z} that analyzes the plasma dynamics under the influence of the external discharge circuit. The results confirm that no pinching occurs by using the same plasma parameters of the ABP.

\begin{figure}[h]
\centering
\includegraphics[width=85mm]{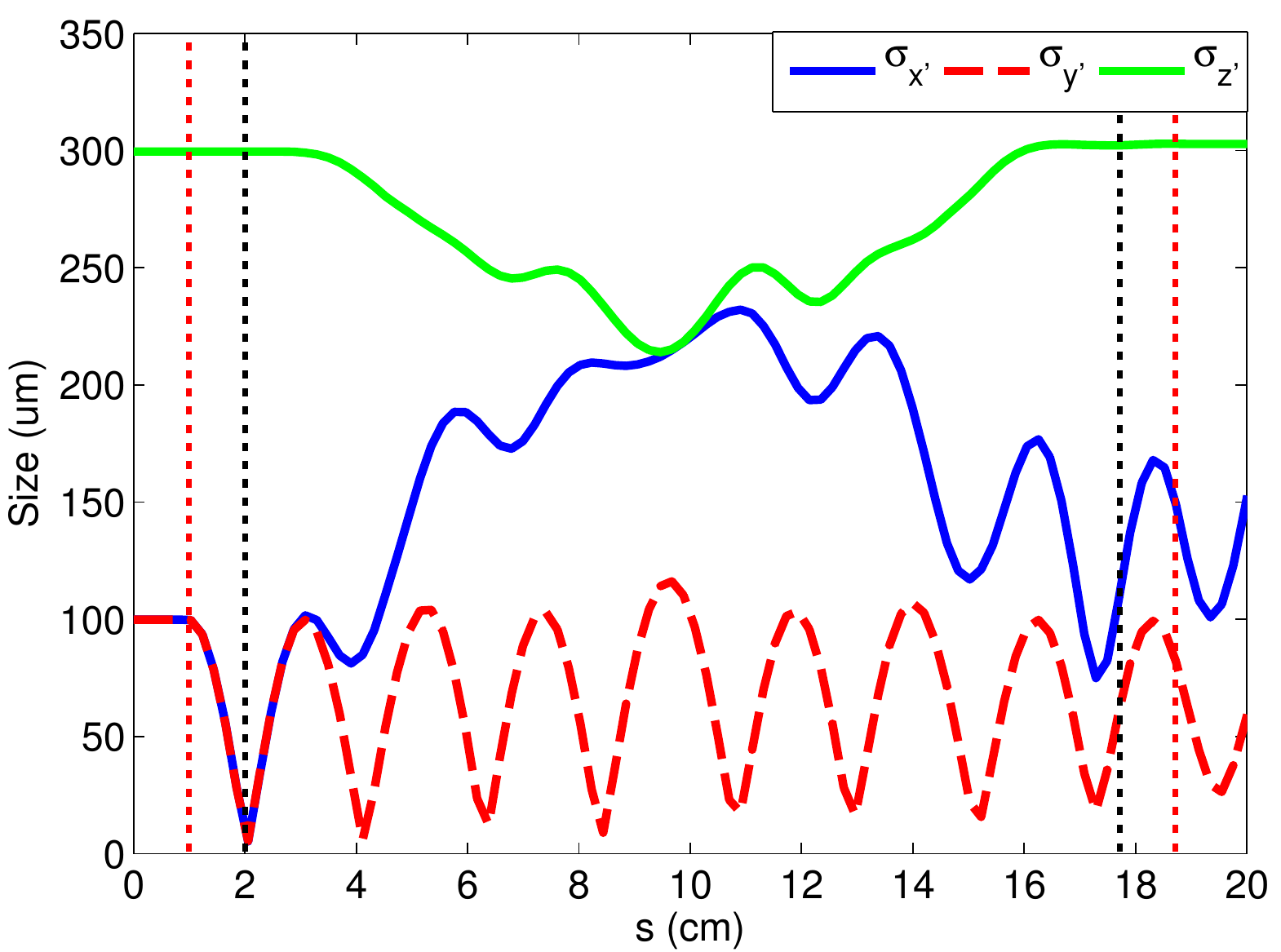}
\caption{Bunch envelopes as a function of the longitudinal position \textit{s}. The transverse sizes are labelled as $\sigma_{x',y'}$ (solid blue and dashed red lines). The bunch length is denoted by $\sigma_{z'}$ (green line). As in Fig.~\ref{traj}, the red (black) dotted lines show the capillary channel (without) including the two straight sections.}
\label{envelopes}
\end{figure}

The magnetic field of the ABP in the full three-dimensional space is then computed with the commercial code CST~STUDIO~\cite{studio2008cst}, that allows to produce the field map along the curved capillary geometry by using the electrical conductivity $\sigma_e (r)$ previously obtained.
Fig.~\ref{FieldMapLinear10_cm} shows the resulting intensity profile along the $y=0$ plane. Two 1~cm long straight sections have been included before and after the curved capillary.

To test the particle deflection, we imported the 3D field map in GPT. The reference electron beam, consisting of $10^6$ macro-particles, is assumed to be injected from the left of Fig.~\ref{FieldMapLinear10_cm} at $x=y=z=0$.
The simulated macro-particle trajectories are reported in Fig.~\ref{traj}. An initial (final) drift of 1~(2)~cm is considered upstream (downstream) the capillary channel. The red (black) dotted lines show the overall path (without) including the two straight sections.
The plot highlights that all the particles are properly bent and follow the curved path along the capillary.
The beam envelopes are shown in Fig.~\ref{envelopes}. Here we refer to the envelopes $\sigma_{x',y',z'}$, i.e. the ones calculated in the beam co-moving system $x',y',z'$. The relative position of the beam along its path is denoted by the  longitudinal coordinate $s$.
One can see that the beam undergoes several betatron oscillations during its motion. These are evident by looking at the evolution of $\sigma_y$ since such a plane is not affected by the deflection. Regarding the $\sigma_x$ envelope, it exactly resembles $\sigma_y$ up to the end of the straight capillary section (approximately at $s=3$~cm), then it starts to follows a different behavior on the plane of deflection.

An interesting feature pointed out by Fig.~\ref{envelopes} is the evolution of the bunch length $\sigma_z$. What comes out is that such a quantity is effectively preserved (within $1\%$ of its initial value) at the end of the bending, thus no elongation (or longitudinal dispersion) of the bunch is observed as in the case of conventional bending magnets.
It is due to the fact that, regardless of their energy, all the particles follow approximately the same path. Indeed, by considering that the focusing provided by the azimuthal magnetic field grows with radial distance from the capillary axis, larger (lower) is the particle energy larger (lower) would be the offset with respect to the reference path since $\rho\propto E$. As a consequence a stronger (weaker) kick is produced and the resulting trajectories are almost independent on the particle energy. 

\begin{figure}[h]
\centering
\includegraphics[width=85mm]{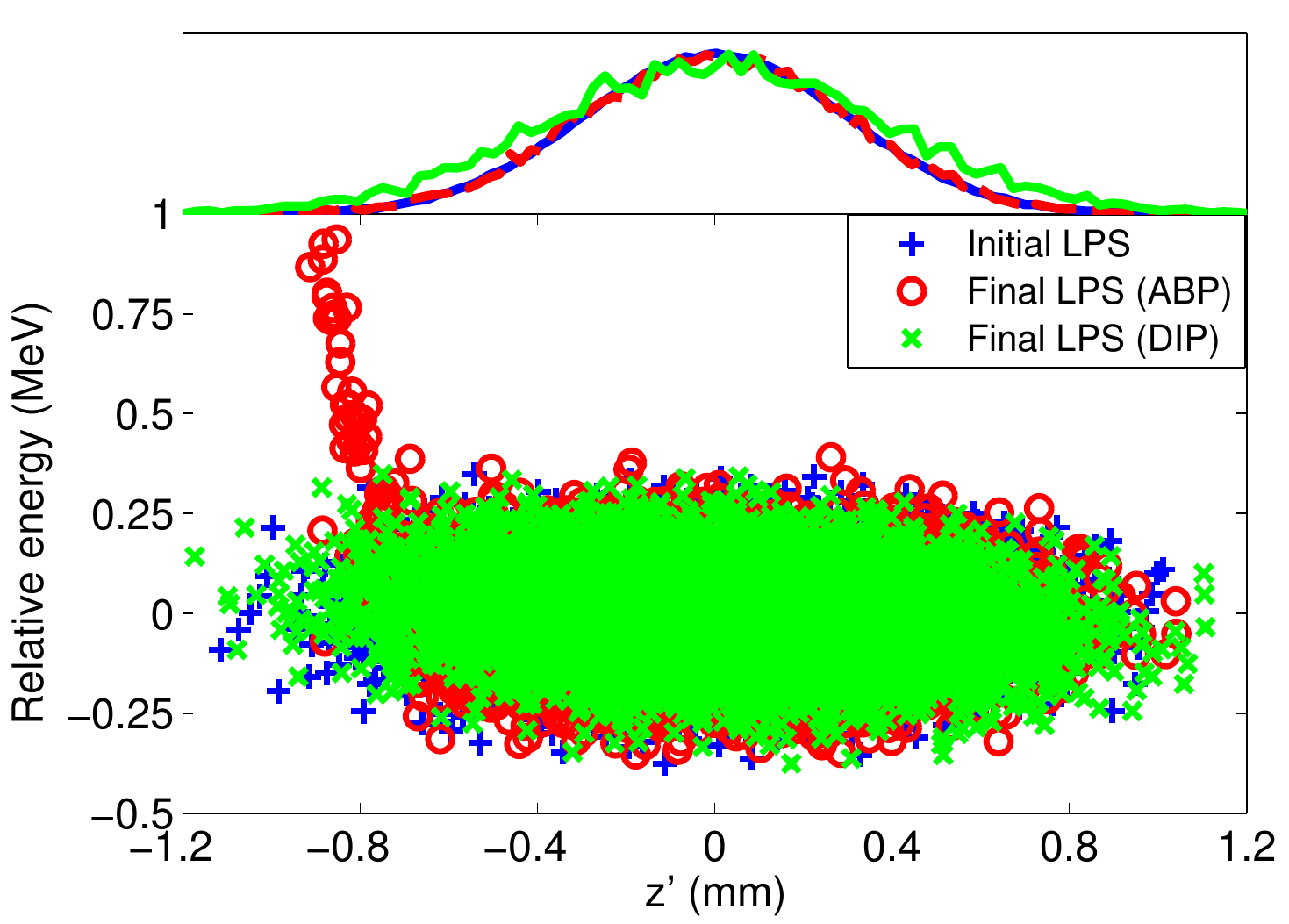}
\caption{Longitudinal phase space (LPS) at the entrance ($\color{blue}+$) and exit ($\color{red}\circ$) of the ABP. The \textit{y}-axis shows the single particle energies with respect to the mean bunch energy (100~MeV). The resulting LPS at the exit of a conventional bending dipole ($\color{green}\times$) is also reported. On top the resulting histogram of the bunch longitudinal profile for the three cases is shown.}
\label{LPS_if}
\end{figure}

Fig.~\ref{LPS_if} compares the initial ($\color{blue}+$) and final ($\color{red}\circ$) beam longitudinal phase space (LPS) computed by Architect.
At the end of the ABP a small fraction (less than $0.1\%$) of particles in the tail is weakly accelerated by the induced plasma wakefield but the main core of the bunch actually remains unaffected and the overall elongation is of the order of $0.4\%$. 
In the same plot we have also reported the resulting LPS at the exit of a conventional bending dipole ($\color{green}\times$), modeled as a sector magnet with same radius of curvature $\rho$ and gap equal to $2 R_c$. In this case the bunch acquires an energy chirp and its elongation is more pronounced (about $8\%$).


\begin{figure}[h]
\centering
\begin{overpic}[height=0.425\textwidth]{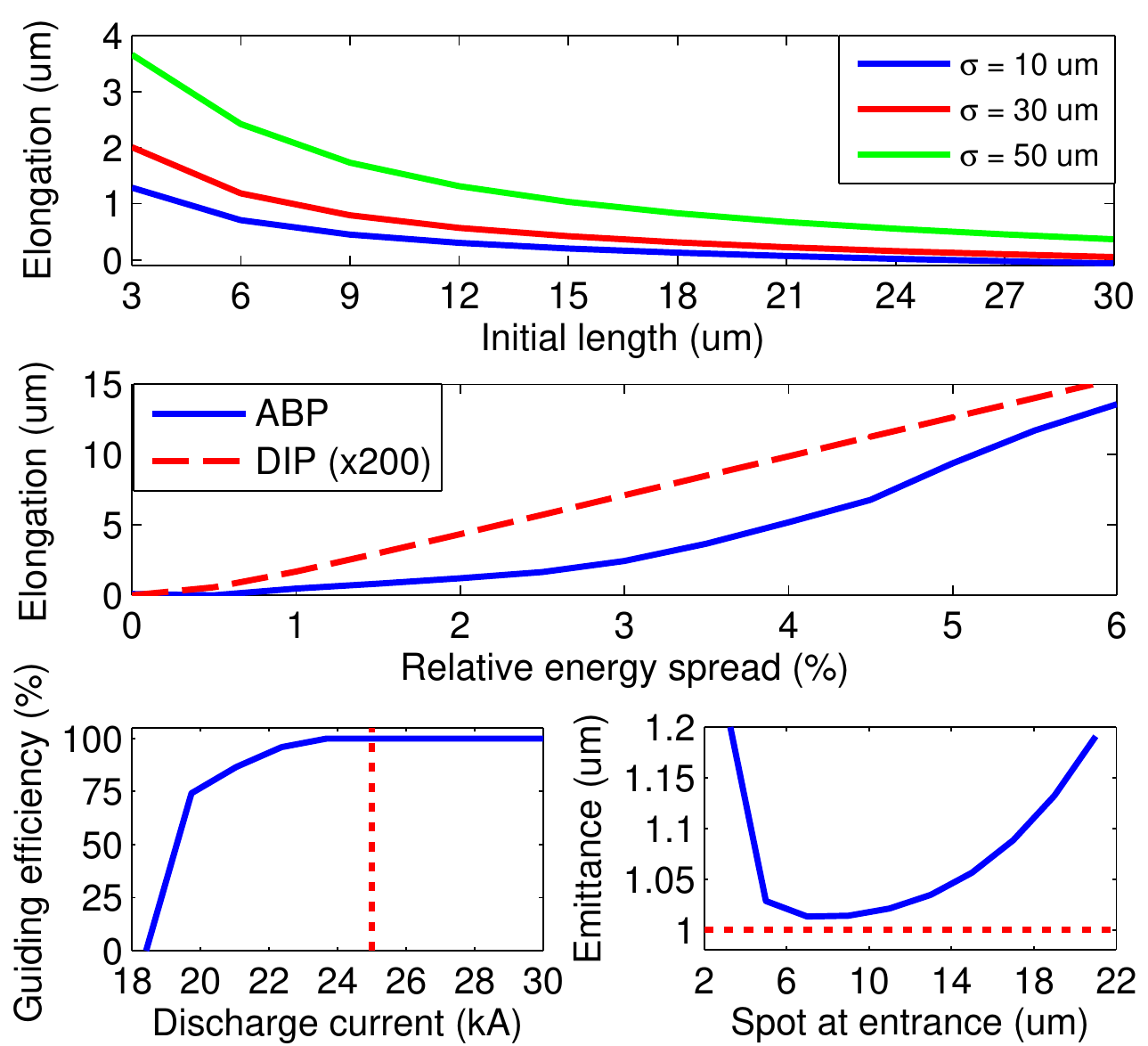}
\put(13,80){\color{black}\textbf{a}}
\put(13,43){\color{black}\textbf{b}}
\put(13,23){\color{black}\textbf{c}}
\put(66,23){\color{black}\textbf{d}}
\end{overpic}
\caption{(a) Bunch elongation as a function of the initial bunch length for several spot sizes at ABP entrance. (b) Elongation calculated at different energy spreads for the reference beam ($\sigma_z = 300~\mu m$) at the end of the ABP (blue line). The red dashed line refers to a conventional bending dipole. Its real values must be multiplied by a factor of 200. (c) Guiding efficiency as a function of the applied discharge current. At lower values a larger fraction of particles is lost. The red dashed line represents the discharge current used for all the calculations considered so far. (d) Resulting emittance at the end of the transport as a function of the initial spot size. The red dashed line indicates the initial emittance ($\epsilon_n=1~\mu m$).}
\label{subplot_MR}
\end{figure}
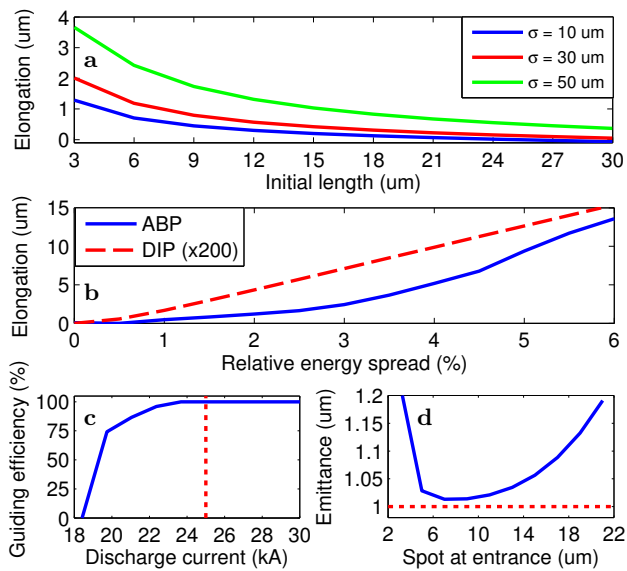


A parametric study of the main features of a possible ABP device is summarized in Fig.~\ref{subplot_MR}.
The effectiveness in preserving the length of even ultra-short bunches (down to $3~\mu m$ or, equivalently, 10~fs) is shown in Fig.~\ref{subplot_MR}(a) for several values of the bunch spot size at the ABP entrance.
One can see that such a quantity is better preserved for small aspect ratio beams, i.e. when $\sigma_{x,y}/\sigma_z\lesssim 1$.
A similar study is reported in Fig.~\ref{subplot_MR}(b) where we evaluated that also for very large energy spreads (up to $6\%$) the elongation is contained within only $5\%$. The same plot highlights that, for a conventional bending dipole, the same values of energy spread would result in an overall elongation approximately 200 times larger.
The ability to guide the beam for different current discharges flowing through the capillary is shown in Fig.~\ref{subplot_MR}(c). One can see that for the reference 100~MeV beam a minimum current $I_D\approx 24$~kA is needed to properly bend all the particles. At lower values the guiding is inefficient and a larger fraction of particles is lost during the transport. On the contrary, at larger currents the guiding is always guaranteed since the resulting magnetic field is large enough to keep all the particles close to the reference trajectory.
Finally, in Fig.~\ref{subplot_MR}(d) we have evaluated the resulting increase of the beam emittance as a function of the spot size at the capillary entrance. It results that the beam emittance is preserved when $\sigma_{x,y}\ll R_c$, which ensures that the magnetic field (as the one calculated in Fig.~\ref{calcProfiles}) is almost linear along the bunch radial profile. 
For the reference beam, instead, we find a dramatic emittance degradation (up to $40~\mu m$) due to its large initial spot size.
The best scenario is obtained when the beam is matched with the focusing channel. By assuming a linear focusing force $F(r)=k^2 r$ (with $k^2= \mu_0 q_e \beta c I_D/2\pi R_c^2$) acting at distance $r$ from the axis, the beam envelope equation is
\begin{equation}
\sigma_{x,y}^{''} + k_n^2 \sigma_{x,y} = {\epsilon_n^2\over \sigma_{x,y}^3}~,
\end{equation}
where $k_n = k/\sqrt{\gamma m_e c^2}$ is the normalized lens focusing strength, $\gamma$ the relativistic Lorentz factor and $m_e$ the electron rest mass.
The equilibrium solution is associated to a betatron oscillation of amplitude $\beta_{eq}=1/k_n$ and, thus, to a matched spot size equal to $\sigma_{eq}=\sqrt{\beta_{eq} \epsilon_n /\gamma}\approx 7~\mu m$ for the beam parameters described so far. For such a value the emittance minimum in Fig.~\ref{subplot_MR}(d) is obtained.

The ABP deflection capability can be scaled to even higher beam energies (and/or larger deflection angles) by changing its main parameters (radius of curvature, discharge current and capillary radius). Here we have demonstrated how to obtain the most compact structure able to bend a 100~MeV electron beam. For other cases, e.g. when dealing with larger energies, the radius of curvature has to be increased (since $\rho\propto E$) by employing longer capillaries and avoiding the use of too large discharge currents that can induce plasma pinch and other nonlinear effects.




In conclusion, we have presented a new device based on a capillary-discharge waveguide able to deflect particles at large angles. 
The theoretical treatment is discussed and supported by numerical simulations showing that the guiding, under certain conditions, preserves both the beam emittance and longitudinal phase space.
The latter feature, in particular, would be particularly useful in accelerator facilities. If the beam has to be translated in a different beamline, a system consisting of (at least two) consecutive bending magnets separated by dispersion-matching focusing optics (e.g. quadrupoles and sextupoles) has to be adopted to preserve (or compress) the bunch duration~\cite{england2005sextupole,eos_jitter}. On the contrary, by means of the ABP even ultra-short beams can be transported without requiring additional optics.
We have also demonstrated its capability in guiding particle beams with large energy spread as, for instance, the ones coming from plasma-based accelerators~\cite{faure2004laser,geddes2004high,litos2014high}. In this case the transport up to a specific location represents a tricky task to accomplish if conventional magnetic optics (strongly affected by chromatic effects) is employed. With the ABP, instead, the overall elongation would be contained within only few percents.
If compared to the state of the art of current technology, based on superconducting magnets operating at cryogenic temperatures, its practical implementation would be simpler and more affordable in terms of costs. 
The ABP might thus represent an affordable solution to develop more compact beamlines and, in general, to bend and guide charged particle beams. 




\begin{acknowledgments}
This work has been partially supported by the EU Commission in the Seventh Framework Program, Grant Agreement 312453-EuCARD-2, the European Union Horizon 2020 research and innovation program, Grant Agreement No. 653782 (EuPRAXIA), and the Italian Research Minister in the framework of FIRB - Fondo per gli Investimenti della Ricerca di Base, Project nr. RBFR12NK5K.
\end{acknowledgments}

\bibliography{biblio}

\begin{thebibliography}{42}%
\makeatletter
\providecommand \@ifxundefined [1]{%
 \@ifx{#1\undefined}
}%
\providecommand \@ifnum [1]{%
 \ifnum #1\expandafter \@firstoftwo
 \else \expandafter \@secondoftwo
 \fi
}%
\providecommand \@ifx [1]{%
 \ifx #1\expandafter \@firstoftwo
 \else \expandafter \@secondoftwo
 \fi
}%
\providecommand \natexlab [1]{#1}%
\providecommand \enquote  [1]{``#1''}%
\providecommand \bibnamefont  [1]{#1}%
\providecommand \bibfnamefont [1]{#1}%
\providecommand \citenamefont [1]{#1}%
\providecommand \href@noop [0]{\@secondoftwo}%
\providecommand \href [0]{\begingroup \@sanitize@url \@href}%
\providecommand \@href[1]{\@@startlink{#1}\@@href}%
\providecommand \@@href[1]{\endgroup#1\@@endlink}%
\providecommand \@sanitize@url [0]{\catcode `\\12\catcode `\$12\catcode
  `\&12\catcode `\#12\catcode `\^12\catcode `\_12\catcode `\%12\relax}%
\providecommand \@@startlink[1]{}%
\providecommand \@@endlink[0]{}%
\providecommand \url  [0]{\begingroup\@sanitize@url \@url }%
\providecommand \@url [1]{\endgroup\@href {#1}{\urlprefix }}%
\providecommand \urlprefix  [0]{URL }%
\providecommand \Eprint [0]{\href }%
\providecommand \doibase [0]{http://dx.doi.org/}%
\providecommand \selectlanguage [0]{\@gobble}%
\providecommand \bibinfo  [0]{\@secondoftwo}%
\providecommand \bibfield  [0]{\@secondoftwo}%
\providecommand \translation [1]{[#1]}%
\providecommand \BibitemOpen [0]{}%
\providecommand \bibitemStop [0]{}%
\providecommand \bibitemNoStop [0]{.\EOS\space}%
\providecommand \EOS [0]{\spacefactor3000\relax}%
\providecommand \BibitemShut  [1]{\csname bibitem#1\endcsname}%
\let\auto@bib@innerbib\@empty
\bibitem [{\citenamefont {Reiser}(2008)}]{reiser}%
  \BibitemOpen
  \bibfield  {author} {\bibinfo {author} {\bibfnamefont {M.}~\bibnamefont
  {Reiser}},\ }\href@noop {} {\emph {\bibinfo {title} {Theory and design of
  charged particle beams}}}\ (\bibinfo  {publisher} {John Wiley \& Sons},\
  \bibinfo {year} {2008})\BibitemShut {NoStop}%
\bibitem [{\citenamefont {Bryant}\ and\ \citenamefont
  {Johnsen}(2005)}]{bryant2005principles}%
  \BibitemOpen
  \bibfield  {author} {\bibinfo {author} {\bibfnamefont {P.~J.}\ \bibnamefont
  {Bryant}}\ and\ \bibinfo {author} {\bibfnamefont {K.}~\bibnamefont
  {Johnsen}},\ }\href@noop {} {\emph {\bibinfo {title} {The principles of
  circular accelerators and storage rings}}}\ (\bibinfo  {publisher} {Cambridge
  University Press},\ \bibinfo {year} {2005})\BibitemShut {NoStop}%
\bibitem [{\citenamefont {Leemann}\ \emph {et~al.}(2009)\citenamefont
  {Leemann}, \citenamefont {Andersson}, \citenamefont {Eriksson}, \citenamefont
  {Lindgren}, \citenamefont {Wall{\'e}n}, \citenamefont {Bengtsson},\ and\
  \citenamefont {Streun}}]{leemann2009beam}%
  \BibitemOpen
  \bibfield  {author} {\bibinfo {author} {\bibfnamefont {S.}~\bibnamefont
  {Leemann}}, \bibinfo {author} {\bibfnamefont {{\AA}.}~\bibnamefont
  {Andersson}}, \bibinfo {author} {\bibfnamefont {M.}~\bibnamefont {Eriksson}},
  \bibinfo {author} {\bibfnamefont {L.-J.}\ \bibnamefont {Lindgren}}, \bibinfo
  {author} {\bibfnamefont {E.}~\bibnamefont {Wall{\'e}n}}, \bibinfo {author}
  {\bibfnamefont {J.}~\bibnamefont {Bengtsson}}, \ and\ \bibinfo {author}
  {\bibfnamefont {A.}~\bibnamefont {Streun}},\ }\href@noop {} {\bibfield
  {journal} {\bibinfo  {journal} {Physical Review Special Topics-Accelerators
  and Beams}\ }\textbf {\bibinfo {volume} {12}},\ \bibinfo {pages} {120701}
  (\bibinfo {year} {2009})}\BibitemShut {NoStop}%
\bibitem [{\citenamefont {Sokolov}\ and\ \citenamefont
  {Ternov}(1966)}]{sokolov1966synchrotron}%
  \BibitemOpen
  \bibfield  {author} {\bibinfo {author} {\bibfnamefont {A.~A.}\ \bibnamefont
  {Sokolov}}\ and\ \bibinfo {author} {\bibfnamefont {I.~M.}\ \bibnamefont
  {Ternov}},\ }\href@noop {} {\bibfield  {journal} {\bibinfo  {journal}
  {Akademia Nauk SSSR, Moskovskoie Obshchestvo Ispytatelei prirody. Sektsia
  Fiziki. Sinkhrotron Radiation, Nauka Eds., Moscow, 1966 (Russian title:
  Sinkhrotronnoie izluchenie), 228 pp.}\ } (\bibinfo {year}
  {1966})}\BibitemShut {NoStop}%
\bibitem [{\citenamefont {Helliwell}(1998)}]{helliwell1998synchrotron}%
  \BibitemOpen
  \bibfield  {author} {\bibinfo {author} {\bibfnamefont {J.~R.}\ \bibnamefont
  {Helliwell}},\ }\href@noop {} {\bibfield  {journal} {\bibinfo  {journal}
  {Nature Structural \& Molecular Biology}\ }\textbf {\bibinfo {volume} {5}},\
  \bibinfo {pages} {614} (\bibinfo {year} {1998})}\BibitemShut {NoStop}%
\bibitem [{\citenamefont {Brown}\ and\ \citenamefont
  {Sturchio}(2002)}]{brown2002overview}%
  \BibitemOpen
  \bibfield  {author} {\bibinfo {author} {\bibfnamefont {G.~E.}\ \bibnamefont
  {Brown}}\ and\ \bibinfo {author} {\bibfnamefont {N.~C.}\ \bibnamefont
  {Sturchio}},\ }\href@noop {} {\bibfield  {journal} {\bibinfo  {journal}
  {Reviews in Mineralogy and Geochemistry}\ }\textbf {\bibinfo {volume} {49}},\
  \bibinfo {pages} {1} (\bibinfo {year} {2002})}\BibitemShut {NoStop}%
\bibitem [{\citenamefont {Weik}\ \emph {et~al.}(2000)\citenamefont {Weik},
  \citenamefont {Ravelli}, \citenamefont {Kryger}, \citenamefont {McSweeney},
  \citenamefont {Raves}, \citenamefont {Harel}, \citenamefont {Gros},
  \citenamefont {Silman}, \citenamefont {Kroon},\ and\ \citenamefont
  {Sussman}}]{weik2000specific}%
  \BibitemOpen
  \bibfield  {author} {\bibinfo {author} {\bibfnamefont {M.}~\bibnamefont
  {Weik}}, \bibinfo {author} {\bibfnamefont {R.~B.}\ \bibnamefont {Ravelli}},
  \bibinfo {author} {\bibfnamefont {G.}~\bibnamefont {Kryger}}, \bibinfo
  {author} {\bibfnamefont {S.}~\bibnamefont {McSweeney}}, \bibinfo {author}
  {\bibfnamefont {M.~L.}\ \bibnamefont {Raves}}, \bibinfo {author}
  {\bibfnamefont {M.}~\bibnamefont {Harel}}, \bibinfo {author} {\bibfnamefont
  {P.}~\bibnamefont {Gros}}, \bibinfo {author} {\bibfnamefont {I.}~\bibnamefont
  {Silman}}, \bibinfo {author} {\bibfnamefont {J.}~\bibnamefont {Kroon}}, \
  and\ \bibinfo {author} {\bibfnamefont {J.~L.}\ \bibnamefont {Sussman}},\
  }\href@noop {} {\bibfield  {journal} {\bibinfo  {journal} {Proceedings of the
  National Academy of Sciences}\ }\textbf {\bibinfo {volume} {97}},\ \bibinfo
  {pages} {623} (\bibinfo {year} {2000})}\BibitemShut {NoStop}%
\bibitem [{\citenamefont {Lewis}(1997)}]{lewis1997medical}%
  \BibitemOpen
  \bibfield  {author} {\bibinfo {author} {\bibfnamefont {R.}~\bibnamefont
  {Lewis}},\ }\href@noop {} {\bibfield  {journal} {\bibinfo  {journal} {Physics
  in medicine and biology}\ }\textbf {\bibinfo {volume} {42}},\ \bibinfo
  {pages} {1213} (\bibinfo {year} {1997})}\BibitemShut {NoStop}%
\bibitem [{\citenamefont {Suortti}\ and\ \citenamefont
  {Thomlinson}(2003)}]{suortti2003medical}%
  \BibitemOpen
  \bibfield  {author} {\bibinfo {author} {\bibfnamefont {P.}~\bibnamefont
  {Suortti}}\ and\ \bibinfo {author} {\bibfnamefont {W.}~\bibnamefont
  {Thomlinson}},\ }\href@noop {} {\bibfield  {journal} {\bibinfo  {journal}
  {Physics in medicine and biology}\ }\textbf {\bibinfo {volume} {48}},\
  \bibinfo {pages} {R1} (\bibinfo {year} {2003})}\BibitemShut {NoStop}%
\bibitem [{\citenamefont {Gemmell}(1974)}]{gemmell1974channeling}%
  \BibitemOpen
  \bibfield  {author} {\bibinfo {author} {\bibfnamefont {D.~S.}\ \bibnamefont
  {Gemmell}},\ }\href@noop {} {\bibfield  {journal} {\bibinfo  {journal}
  {Reviews of Modern Physics}\ }\textbf {\bibinfo {volume} {46}},\ \bibinfo
  {pages} {129} (\bibinfo {year} {1974})}\BibitemShut {NoStop}%
\bibitem [{\citenamefont {Dabagov}\ \emph {et~al.}(2015)\citenamefont
  {Dabagov}, \citenamefont {Dik},\ and\ \citenamefont
  {Frolov}}]{dabagov2015channeling}%
  \BibitemOpen
  \bibfield  {author} {\bibinfo {author} {\bibfnamefont {S.~B.}\ \bibnamefont
  {Dabagov}}, \bibinfo {author} {\bibfnamefont {A.~V.}\ \bibnamefont {Dik}}, \
  and\ \bibinfo {author} {\bibfnamefont {E.~N.}\ \bibnamefont {Frolov}},\
  }\href@noop {} {\bibfield  {journal} {\bibinfo  {journal} {Physical Review
  Special Topics-Accelerators and Beams}\ }\textbf {\bibinfo {volume} {18}},\
  \bibinfo {pages} {064002} (\bibinfo {year} {2015})}\BibitemShut {NoStop}%
\bibitem [{\citenamefont {Tomita}\ and\ \citenamefont
  {Murakami}(2003)}]{tomita2003high}%
  \BibitemOpen
  \bibfield  {author} {\bibinfo {author} {\bibfnamefont {M.}~\bibnamefont
  {Tomita}}\ and\ \bibinfo {author} {\bibfnamefont {M.}~\bibnamefont
  {Murakami}},\ }\href@noop {} {\bibfield  {journal} {\bibinfo  {journal}
  {Nature}\ }\textbf {\bibinfo {volume} {421}},\ \bibinfo {pages} {517}
  (\bibinfo {year} {2003})}\BibitemShut {NoStop}%
\bibitem [{\citenamefont {Todesco}\ \emph {et~al.}(2004)\citenamefont
  {Todesco}, \citenamefont {Bellesia}, \citenamefont {Bottura}, \citenamefont
  {Devred}, \citenamefont {Remondino}, \citenamefont {Pauletta}, \citenamefont
  {Sanfilippo}, \citenamefont {Scandale}, \citenamefont {Vollinger},\ and\
  \citenamefont {Wildner}}]{todesco2004steering}%
  \BibitemOpen
  \bibfield  {author} {\bibinfo {author} {\bibfnamefont {E.}~\bibnamefont
  {Todesco}}, \bibinfo {author} {\bibfnamefont {B.}~\bibnamefont {Bellesia}},
  \bibinfo {author} {\bibfnamefont {L.}~\bibnamefont {Bottura}}, \bibinfo
  {author} {\bibfnamefont {A.}~\bibnamefont {Devred}}, \bibinfo {author}
  {\bibfnamefont {V.}~\bibnamefont {Remondino}}, \bibinfo {author}
  {\bibfnamefont {S.}~\bibnamefont {Pauletta}}, \bibinfo {author}
  {\bibfnamefont {S.}~\bibnamefont {Sanfilippo}}, \bibinfo {author}
  {\bibfnamefont {W.}~\bibnamefont {Scandale}}, \bibinfo {author}
  {\bibfnamefont {C.}~\bibnamefont {Vollinger}}, \ and\ \bibinfo {author}
  {\bibfnamefont {E.}~\bibnamefont {Wildner}},\ }\href@noop {} {\bibfield
  {journal} {\bibinfo  {journal} {IEEE transactions on applied
  superconductivity}\ }\textbf {\bibinfo {volume} {14}},\ \bibinfo {pages}
  {177} (\bibinfo {year} {2004})}\BibitemShut {NoStop}%
\bibitem [{\citenamefont {Evans}\ and\ \citenamefont
  {Bryant}(2008)}]{evans2008lhc}%
  \BibitemOpen
  \bibfield  {author} {\bibinfo {author} {\bibfnamefont {L.}~\bibnamefont
  {Evans}}\ and\ \bibinfo {author} {\bibfnamefont {P.}~\bibnamefont {Bryant}},\
  }\href@noop {} {\bibfield  {journal} {\bibinfo  {journal} {Journal of
  instrumentation}\ }\textbf {\bibinfo {volume} {3}},\ \bibinfo {pages}
  {S08001} (\bibinfo {year} {2008})}\BibitemShut {NoStop}%
\bibitem [{\citenamefont {{Tajima}}\ and\ \citenamefont
  {{Dawson}}(1979)}]{1979PhRvL..43..267T}%
  \BibitemOpen
  \bibfield  {author} {\bibinfo {author} {\bibfnamefont {T.}~\bibnamefont
  {{Tajima}}}\ and\ \bibinfo {author} {\bibfnamefont {J.~M.}\ \bibnamefont
  {{Dawson}}},\ }\href {\doibase 10.1103/PhysRevLett.43.267} {\bibfield
  {journal} {\bibinfo  {journal} {Physical Review Letters}\ }\textbf {\bibinfo
  {volume} {43}},\ \bibinfo {pages} {267} (\bibinfo {year} {1979})}\BibitemShut
  {NoStop}%
\bibitem [{\citenamefont {Rocca}\ \emph {et~al.}(1993)\citenamefont {Rocca},
  \citenamefont {Cortazar}, \citenamefont {Szapiro}, \citenamefont {Floyd},\
  and\ \citenamefont {Tomasel}}]{rocca1993fast}%
  \BibitemOpen
  \bibfield  {author} {\bibinfo {author} {\bibfnamefont {J.}~\bibnamefont
  {Rocca}}, \bibinfo {author} {\bibfnamefont {O.}~\bibnamefont {Cortazar}},
  \bibinfo {author} {\bibfnamefont {B.}~\bibnamefont {Szapiro}}, \bibinfo
  {author} {\bibfnamefont {K.}~\bibnamefont {Floyd}}, \ and\ \bibinfo {author}
  {\bibfnamefont {F.}~\bibnamefont {Tomasel}},\ }\href@noop {} {\bibfield
  {journal} {\bibinfo  {journal} {Physical Review E}\ }\textbf {\bibinfo
  {volume} {47}},\ \bibinfo {pages} {1299} (\bibinfo {year}
  {1993})}\BibitemShut {NoStop}%
\bibitem [{\citenamefont {Rocca}\ \emph {et~al.}(1996)\citenamefont {Rocca},
  \citenamefont {Clark}, \citenamefont {Chilla},\ and\ \citenamefont
  {Shlyaptsev}}]{rocca1996energy}%
  \BibitemOpen
  \bibfield  {author} {\bibinfo {author} {\bibfnamefont {J.~J.}\ \bibnamefont
  {Rocca}}, \bibinfo {author} {\bibfnamefont {D.}~\bibnamefont {Clark}},
  \bibinfo {author} {\bibfnamefont {J.}~\bibnamefont {Chilla}}, \ and\ \bibinfo
  {author} {\bibfnamefont {V.}~\bibnamefont {Shlyaptsev}},\ }\href@noop {}
  {\bibfield  {journal} {\bibinfo  {journal} {Physical review letters}\
  }\textbf {\bibinfo {volume} {77}},\ \bibinfo {pages} {1476} (\bibinfo {year}
  {1996})}\BibitemShut {NoStop}%
\bibitem [{\citenamefont {Tomasel}\ \emph {et~al.}(1997)\citenamefont
  {Tomasel}, \citenamefont {Rocca}, \citenamefont {Shlyaptsev},\ and\
  \citenamefont {Macchietto}}]{tomasel1997lasing}%
  \BibitemOpen
  \bibfield  {author} {\bibinfo {author} {\bibfnamefont {F.}~\bibnamefont
  {Tomasel}}, \bibinfo {author} {\bibfnamefont {J.}~\bibnamefont {Rocca}},
  \bibinfo {author} {\bibfnamefont {V.}~\bibnamefont {Shlyaptsev}}, \ and\
  \bibinfo {author} {\bibfnamefont {C.}~\bibnamefont {Macchietto}},\
  }\href@noop {} {\bibfield  {journal} {\bibinfo  {journal} {Physical Review
  A}\ }\textbf {\bibinfo {volume} {55}},\ \bibinfo {pages} {1437} (\bibinfo
  {year} {1997})}\BibitemShut {NoStop}%
\bibitem [{\citenamefont {Hosokai}\ \emph {et~al.}(2000)\citenamefont
  {Hosokai}, \citenamefont {Kando}, \citenamefont {Dewa}, \citenamefont
  {Kotaki}, \citenamefont {Kondo}, \citenamefont {Hasegawa}, \citenamefont
  {Nakajima},\ and\ \citenamefont {Horioka}}]{hosokai2000optical}%
  \BibitemOpen
  \bibfield  {author} {\bibinfo {author} {\bibfnamefont {T.}~\bibnamefont
  {Hosokai}}, \bibinfo {author} {\bibfnamefont {M.}~\bibnamefont {Kando}},
  \bibinfo {author} {\bibfnamefont {H.}~\bibnamefont {Dewa}}, \bibinfo {author}
  {\bibfnamefont {H.}~\bibnamefont {Kotaki}}, \bibinfo {author} {\bibfnamefont
  {S.}~\bibnamefont {Kondo}}, \bibinfo {author} {\bibfnamefont
  {N.}~\bibnamefont {Hasegawa}}, \bibinfo {author} {\bibfnamefont
  {K.}~\bibnamefont {Nakajima}}, \ and\ \bibinfo {author} {\bibfnamefont
  {K.}~\bibnamefont {Horioka}},\ }\href@noop {} {\bibfield  {journal} {\bibinfo
   {journal} {Optics Letters}\ }\textbf {\bibinfo {volume} {25}},\ \bibinfo
  {pages} {10} (\bibinfo {year} {2000})}\BibitemShut {NoStop}%
\bibitem [{\citenamefont {Leemans}\ \emph {et~al.}(2006)\citenamefont
  {Leemans}, \citenamefont {Nagler}, \citenamefont {Gonsalves}, \citenamefont
  {T{\'o}th}, \citenamefont {Nakamura}, \citenamefont {Geddes}, \citenamefont
  {Esarey}, \citenamefont {Schroeder},\ and\ \citenamefont
  {Hooker}}]{leemans2006gev}%
  \BibitemOpen
  \bibfield  {author} {\bibinfo {author} {\bibfnamefont {W.}~\bibnamefont
  {Leemans}}, \bibinfo {author} {\bibfnamefont {B.}~\bibnamefont {Nagler}},
  \bibinfo {author} {\bibfnamefont {A.}~\bibnamefont {Gonsalves}}, \bibinfo
  {author} {\bibfnamefont {C.}~\bibnamefont {T{\'o}th}}, \bibinfo {author}
  {\bibfnamefont {K.}~\bibnamefont {Nakamura}}, \bibinfo {author}
  {\bibfnamefont {C.}~\bibnamefont {Geddes}}, \bibinfo {author} {\bibfnamefont
  {E.}~\bibnamefont {Esarey}}, \bibinfo {author} {\bibfnamefont
  {C.}~\bibnamefont {Schroeder}}, \ and\ \bibinfo {author} {\bibfnamefont
  {S.}~\bibnamefont {Hooker}},\ }\href@noop {} {\bibfield  {journal} {\bibinfo
  {journal} {Nature physics}\ }\textbf {\bibinfo {volume} {2}},\ \bibinfo
  {pages} {696} (\bibinfo {year} {2006})}\BibitemShut {NoStop}%
\bibitem [{\citenamefont {{Blumenfeld}}\ \emph {et~al.}(2007)\citenamefont
  {{Blumenfeld}}, \citenamefont {{Clayton}}, \citenamefont {{Decker}},
  \citenamefont {{Hogan}}, \citenamefont {{Huang}}, \citenamefont
  {{Ischebeck}}, \citenamefont {{Iverson}}, \citenamefont {{Joshi}},
  \citenamefont {{Katsouleas}}, \citenamefont {{Kirby}}, \citenamefont {{Lu}},
  \citenamefont {{Marsh}}, \citenamefont {{Mori}}, \citenamefont {{Muggli}},
  \citenamefont {{Oz}}, \citenamefont {{Siemann}}, \citenamefont {{Walz}},\
  and\ \citenamefont {{Zhou}}}]{2007Natur.445..741B}%
  \BibitemOpen
  \bibfield  {author} {\bibinfo {author} {\bibfnamefont {I.}~\bibnamefont
  {{Blumenfeld}}}, \bibinfo {author} {\bibfnamefont {C.~E.}\ \bibnamefont
  {{Clayton}}}, \bibinfo {author} {\bibfnamefont {F.-J.}\ \bibnamefont
  {{Decker}}}, \bibinfo {author} {\bibfnamefont {M.~J.}\ \bibnamefont
  {{Hogan}}}, \bibinfo {author} {\bibfnamefont {C.}~\bibnamefont {{Huang}}},
  \bibinfo {author} {\bibfnamefont {R.}~\bibnamefont {{Ischebeck}}}, \bibinfo
  {author} {\bibfnamefont {R.}~\bibnamefont {{Iverson}}}, \bibinfo {author}
  {\bibfnamefont {C.}~\bibnamefont {{Joshi}}}, \bibinfo {author} {\bibfnamefont
  {T.}~\bibnamefont {{Katsouleas}}}, \bibinfo {author} {\bibfnamefont
  {N.}~\bibnamefont {{Kirby}}}, \bibinfo {author} {\bibfnamefont
  {W.}~\bibnamefont {{Lu}}}, \bibinfo {author} {\bibfnamefont {K.~A.}\
  \bibnamefont {{Marsh}}}, \bibinfo {author} {\bibfnamefont {W.~B.}\
  \bibnamefont {{Mori}}}, \bibinfo {author} {\bibfnamefont {P.}~\bibnamefont
  {{Muggli}}}, \bibinfo {author} {\bibfnamefont {E.}~\bibnamefont {{Oz}}},
  \bibinfo {author} {\bibfnamefont {R.~H.}\ \bibnamefont {{Siemann}}}, \bibinfo
  {author} {\bibfnamefont {D.}~\bibnamefont {{Walz}}}, \ and\ \bibinfo {author}
  {\bibfnamefont {M.}~\bibnamefont {{Zhou}}},\ }\href {\doibase
  10.1038/nature05538} {\bibfield  {journal} {\bibinfo  {journal} {Nature}\
  }\textbf {\bibinfo {volume} {445}},\ \bibinfo {pages} {741} (\bibinfo {year}
  {2007})}\BibitemShut {NoStop}%
\bibitem [{\citenamefont {Su}\ \emph {et~al.}(1990)\citenamefont {Su},
  \citenamefont {Katsouleas}, \citenamefont {Dawson},\ and\ \citenamefont
  {Fedele}}]{su1990plasma}%
  \BibitemOpen
  \bibfield  {author} {\bibinfo {author} {\bibfnamefont {J.}~\bibnamefont
  {Su}}, \bibinfo {author} {\bibfnamefont {T.}~\bibnamefont {Katsouleas}},
  \bibinfo {author} {\bibfnamefont {J.}~\bibnamefont {Dawson}}, \ and\ \bibinfo
  {author} {\bibfnamefont {R.}~\bibnamefont {Fedele}},\ }\href@noop {}
  {\bibfield  {journal} {\bibinfo  {journal} {Physical Review A}\ }\textbf
  {\bibinfo {volume} {41}},\ \bibinfo {pages} {3321} (\bibinfo {year}
  {1990})}\BibitemShut {NoStop}%
\bibitem [{\citenamefont {Boggasch}\ \emph {et~al.}(1992)\citenamefont
  {Boggasch}, \citenamefont {Tauschwitz}, \citenamefont {Wahl}, \citenamefont
  {Dietrich}, \citenamefont {Hoffmann}, \citenamefont {Laux}, \citenamefont
  {Stetter},\ and\ \citenamefont {Tkotz}}]{boggasch1992plasma}%
  \BibitemOpen
  \bibfield  {author} {\bibinfo {author} {\bibfnamefont {E.}~\bibnamefont
  {Boggasch}}, \bibinfo {author} {\bibfnamefont {A.}~\bibnamefont
  {Tauschwitz}}, \bibinfo {author} {\bibfnamefont {H.}~\bibnamefont {Wahl}},
  \bibinfo {author} {\bibfnamefont {K.-G.}\ \bibnamefont {Dietrich}}, \bibinfo
  {author} {\bibfnamefont {D.}~\bibnamefont {Hoffmann}}, \bibinfo {author}
  {\bibfnamefont {W.}~\bibnamefont {Laux}}, \bibinfo {author} {\bibfnamefont
  {M.}~\bibnamefont {Stetter}}, \ and\ \bibinfo {author} {\bibfnamefont
  {R.}~\bibnamefont {Tkotz}},\ }\href@noop {} {\bibfield  {journal} {\bibinfo
  {journal} {Applied physics letters}\ }\textbf {\bibinfo {volume} {60}},\
  \bibinfo {pages} {2475} (\bibinfo {year} {1992})}\BibitemShut {NoStop}%
\bibitem [{\citenamefont {Hairapetian}\ \emph {et~al.}(1994)\citenamefont
  {Hairapetian}, \citenamefont {Davis}, \citenamefont {Clayton}, \citenamefont
  {Joshi}, \citenamefont {Hartman}, \citenamefont {Pellegrini},\ and\
  \citenamefont {Katsouleas}}]{hairapetian1994experimental}%
  \BibitemOpen
  \bibfield  {author} {\bibinfo {author} {\bibfnamefont {G.}~\bibnamefont
  {Hairapetian}}, \bibinfo {author} {\bibfnamefont {P.}~\bibnamefont {Davis}},
  \bibinfo {author} {\bibfnamefont {C.}~\bibnamefont {Clayton}}, \bibinfo
  {author} {\bibfnamefont {C.}~\bibnamefont {Joshi}}, \bibinfo {author}
  {\bibfnamefont {S.}~\bibnamefont {Hartman}}, \bibinfo {author} {\bibfnamefont
  {C.}~\bibnamefont {Pellegrini}}, \ and\ \bibinfo {author} {\bibfnamefont
  {T.}~\bibnamefont {Katsouleas}},\ }\href@noop {} {\bibfield  {journal}
  {\bibinfo  {journal} {Physical review letters}\ }\textbf {\bibinfo {volume}
  {72}},\ \bibinfo {pages} {2403} (\bibinfo {year} {1994})}\BibitemShut
  {NoStop}%
\bibitem [{\citenamefont {Van~Tilborg}\ \emph {et~al.}(2015)\citenamefont
  {Van~Tilborg}, \citenamefont {Steinke}, \citenamefont {Geddes}, \citenamefont
  {Matlis}, \citenamefont {Shaw}, \citenamefont {Gonsalves}, \citenamefont
  {Huijts}, \citenamefont {Nakamura}, \citenamefont {Daniels}, \citenamefont
  {Schroeder} \emph {et~al.}}]{van2015active}%
  \BibitemOpen
  \bibfield  {author} {\bibinfo {author} {\bibfnamefont {J.}~\bibnamefont
  {Van~Tilborg}}, \bibinfo {author} {\bibfnamefont {S.}~\bibnamefont
  {Steinke}}, \bibinfo {author} {\bibfnamefont {C.}~\bibnamefont {Geddes}},
  \bibinfo {author} {\bibfnamefont {N.}~\bibnamefont {Matlis}}, \bibinfo
  {author} {\bibfnamefont {B.}~\bibnamefont {Shaw}}, \bibinfo {author}
  {\bibfnamefont {A.}~\bibnamefont {Gonsalves}}, \bibinfo {author}
  {\bibfnamefont {J.}~\bibnamefont {Huijts}}, \bibinfo {author} {\bibfnamefont
  {K.}~\bibnamefont {Nakamura}}, \bibinfo {author} {\bibfnamefont
  {J.}~\bibnamefont {Daniels}}, \bibinfo {author} {\bibfnamefont
  {C.}~\bibnamefont {Schroeder}},  \emph {et~al.},\ }\href@noop {} {\bibfield
  {journal} {\bibinfo  {journal} {Physical review letters}\ }\textbf {\bibinfo
  {volume} {115}},\ \bibinfo {pages} {184802} (\bibinfo {year}
  {2015})}\BibitemShut {NoStop}%
\bibitem [{\citenamefont {Pompili}\ \emph {et~al.}(2017)\citenamefont
  {Pompili}, \citenamefont {Anania}, \citenamefont {Bellaveglia}, \citenamefont
  {Biagioni}, \citenamefont {Bini}, \citenamefont {Bisesto}, \citenamefont
  {Brentegani}, \citenamefont {Castorina}, \citenamefont {Chiadroni},
  \citenamefont {Cianchi} \emph {et~al.}}]{pompili2017experimental}%
  \BibitemOpen
  \bibfield  {author} {\bibinfo {author} {\bibfnamefont {R.}~\bibnamefont
  {Pompili}}, \bibinfo {author} {\bibfnamefont {M.}~\bibnamefont {Anania}},
  \bibinfo {author} {\bibfnamefont {M.}~\bibnamefont {Bellaveglia}}, \bibinfo
  {author} {\bibfnamefont {A.}~\bibnamefont {Biagioni}}, \bibinfo {author}
  {\bibfnamefont {S.}~\bibnamefont {Bini}}, \bibinfo {author} {\bibfnamefont
  {F.}~\bibnamefont {Bisesto}}, \bibinfo {author} {\bibfnamefont
  {E.}~\bibnamefont {Brentegani}}, \bibinfo {author} {\bibfnamefont
  {G.}~\bibnamefont {Castorina}}, \bibinfo {author} {\bibfnamefont
  {E.}~\bibnamefont {Chiadroni}}, \bibinfo {author} {\bibfnamefont
  {A.}~\bibnamefont {Cianchi}},  \emph {et~al.},\ }\href@noop {} {\bibfield
  {journal} {\bibinfo  {journal} {Applied Physics Letters}\ }\textbf {\bibinfo
  {volume} {110}},\ \bibinfo {pages} {104101} (\bibinfo {year}
  {2017})}\BibitemShut {NoStop}%
\bibitem [{\citenamefont {Butler}\ \emph {et~al.}(2002)\citenamefont {Butler},
  \citenamefont {Spence},\ and\ \citenamefont {Hooker}}]{butler2002guiding}%
  \BibitemOpen
  \bibfield  {author} {\bibinfo {author} {\bibfnamefont {A.}~\bibnamefont
  {Butler}}, \bibinfo {author} {\bibfnamefont {D.}~\bibnamefont {Spence}}, \
  and\ \bibinfo {author} {\bibfnamefont {S.~M.}\ \bibnamefont {Hooker}},\
  }\href@noop {} {\bibfield  {journal} {\bibinfo  {journal} {Physical Review
  Letters}\ }\textbf {\bibinfo {volume} {89}},\ \bibinfo {pages} {185003}
  (\bibinfo {year} {2002})}\BibitemShut {NoStop}%
\bibitem [{\citenamefont {Spence}\ \emph {et~al.}(2003)\citenamefont {Spence},
  \citenamefont {Butler},\ and\ \citenamefont {Hooker}}]{spence2003gas}%
  \BibitemOpen
  \bibfield  {author} {\bibinfo {author} {\bibfnamefont {D.}~\bibnamefont
  {Spence}}, \bibinfo {author} {\bibfnamefont {A.}~\bibnamefont {Butler}}, \
  and\ \bibinfo {author} {\bibfnamefont {S.}~\bibnamefont {Hooker}},\
  }\href@noop {} {\bibfield  {journal} {\bibinfo  {journal} {JOSA B}\ }\textbf
  {\bibinfo {volume} {20}},\ \bibinfo {pages} {138} (\bibinfo {year}
  {2003})}\BibitemShut {NoStop}%
\bibitem [{\citenamefont {Ehrlich}\ \emph {et~al.}(1996)\citenamefont
  {Ehrlich}, \citenamefont {Cohen}, \citenamefont {Zigler}, \citenamefont
  {Krall}, \citenamefont {Sprangle},\ and\ \citenamefont
  {Esarey}}]{ehrlich1996guiding}%
  \BibitemOpen
  \bibfield  {author} {\bibinfo {author} {\bibfnamefont {Y.}~\bibnamefont
  {Ehrlich}}, \bibinfo {author} {\bibfnamefont {C.}~\bibnamefont {Cohen}},
  \bibinfo {author} {\bibfnamefont {A.}~\bibnamefont {Zigler}}, \bibinfo
  {author} {\bibfnamefont {J.}~\bibnamefont {Krall}}, \bibinfo {author}
  {\bibfnamefont {P.}~\bibnamefont {Sprangle}}, \ and\ \bibinfo {author}
  {\bibfnamefont {E.}~\bibnamefont {Esarey}},\ }\href@noop {} {\bibfield
  {journal} {\bibinfo  {journal} {Physical review letters}\ }\textbf {\bibinfo
  {volume} {77}},\ \bibinfo {pages} {4186} (\bibinfo {year}
  {1996})}\BibitemShut {NoStop}%
\bibitem [{\citenamefont {Reitsma}\ and\ \citenamefont
  {Jaroszynski}(2008)}]{reitsma2008propagation}%
  \BibitemOpen
  \bibfield  {author} {\bibinfo {author} {\bibfnamefont {A.}~\bibnamefont
  {Reitsma}}\ and\ \bibinfo {author} {\bibfnamefont {D.}~\bibnamefont
  {Jaroszynski}},\ }\href@noop {} {\bibfield  {journal} {\bibinfo  {journal}
  {IEEE Transactions on Plasma Science}\ }\textbf {\bibinfo {volume} {36}},\
  \bibinfo {pages} {1738} (\bibinfo {year} {2008})}\BibitemShut {NoStop}%
\bibitem [{Note1()}]{Note1}%
  \BibitemOpen
  \bibinfo {note} {All the quantities are quoted as root mean square
  (rms).}\BibitemShut {Stop}%
\bibitem [{\citenamefont {De~Loos}\ and\ \citenamefont {Van~der
  Geer}(1996)}]{de1996general}%
  \BibitemOpen
  \bibfield  {author} {\bibinfo {author} {\bibfnamefont {M.}~\bibnamefont
  {De~Loos}}\ and\ \bibinfo {author} {\bibfnamefont {S.}~\bibnamefont {Van~der
  Geer}},\ }in\ \href@noop {} {\emph {\bibinfo {booktitle} {5th European
  Particle Accelerator Conference}}}\ (\bibinfo {year} {1996})\ p.\ \bibinfo
  {pages} {1241}\BibitemShut {NoStop}%
\bibitem [{\citenamefont {Marocchino}\ \emph {et~al.}(2016)\citenamefont
  {Marocchino}, \citenamefont {Massimo}, \citenamefont {Rossi}, \citenamefont
  {Chiadroni},\ and\ \citenamefont {Ferrario}}]{marocchino2016efficient}%
  \BibitemOpen
  \bibfield  {author} {\bibinfo {author} {\bibfnamefont {A.}~\bibnamefont
  {Marocchino}}, \bibinfo {author} {\bibfnamefont {F.}~\bibnamefont {Massimo}},
  \bibinfo {author} {\bibfnamefont {A.}~\bibnamefont {Rossi}}, \bibinfo
  {author} {\bibfnamefont {E.}~\bibnamefont {Chiadroni}}, \ and\ \bibinfo
  {author} {\bibfnamefont {M.}~\bibnamefont {Ferrario}},\ }\href@noop {}
  {\bibfield  {journal} {\bibinfo  {journal} {Nuclear Instruments and Methods
  in Physics Research Section A: Accelerators, Spectrometers, Detectors and
  Associated Equipment}\ }\textbf {\bibinfo {volume} {829}},\ \bibinfo {pages}
  {386} (\bibinfo {year} {2016})}\BibitemShut {NoStop}%
\bibitem [{\citenamefont {Bobrova}\ \emph {et~al.}(2001)\citenamefont
  {Bobrova}, \citenamefont {Esaulov}, \citenamefont {Sakai}, \citenamefont
  {Sasorov}, \citenamefont {Spence}, \citenamefont {Butler}, \citenamefont
  {Hooker},\ and\ \citenamefont {Bulanov}}]{bobrova2001simulations}%
  \BibitemOpen
  \bibfield  {author} {\bibinfo {author} {\bibfnamefont {N.}~\bibnamefont
  {Bobrova}}, \bibinfo {author} {\bibfnamefont {A.}~\bibnamefont {Esaulov}},
  \bibinfo {author} {\bibfnamefont {J.-I.}\ \bibnamefont {Sakai}}, \bibinfo
  {author} {\bibfnamefont {P.}~\bibnamefont {Sasorov}}, \bibinfo {author}
  {\bibfnamefont {D.}~\bibnamefont {Spence}}, \bibinfo {author} {\bibfnamefont
  {A.}~\bibnamefont {Butler}}, \bibinfo {author} {\bibfnamefont
  {S.}~\bibnamefont {Hooker}}, \ and\ \bibinfo {author} {\bibfnamefont
  {S.}~\bibnamefont {Bulanov}},\ }\href@noop {} {\bibfield  {journal} {\bibinfo
   {journal} {Physical Review E}\ }\textbf {\bibinfo {volume} {65}},\ \bibinfo
  {pages} {016407} (\bibinfo {year} {2001})}\BibitemShut {NoStop}%
\bibitem [{\citenamefont {Bennett}(1955)}]{bennett1955self}%
  \BibitemOpen
  \bibfield  {author} {\bibinfo {author} {\bibfnamefont {W.~H.}\ \bibnamefont
  {Bennett}},\ }\href@noop {} {\bibfield  {journal} {\bibinfo  {journal}
  {Physical Review}\ }\textbf {\bibinfo {volume} {98}},\ \bibinfo {pages}
  {1584} (\bibinfo {year} {1955})}\BibitemShut {NoStop}%
\bibitem [{\citenamefont {Vrba}\ and\ \citenamefont
  {Vrbov{\'a}}(2000)}]{vrba2000z}%
  \BibitemOpen
  \bibfield  {author} {\bibinfo {author} {\bibfnamefont {P.}~\bibnamefont
  {Vrba}}\ and\ \bibinfo {author} {\bibfnamefont {M.}~\bibnamefont
  {Vrbov{\'a}}},\ }\href@noop {} {\bibfield  {journal} {\bibinfo  {journal}
  {Contributions to Plasma Physics}\ }\textbf {\bibinfo {volume} {40}},\
  \bibinfo {pages} {581} (\bibinfo {year} {2000})}\BibitemShut {NoStop}%
\bibitem [{\citenamefont {Studio}(2008)}]{studio2008cst}%
  \BibitemOpen
  \bibfield  {author} {\bibinfo {author} {\bibfnamefont {M.}~\bibnamefont
  {Studio}},\ }\href@noop {} {\bibfield  {journal} {\bibinfo  {journal} {Bad
  Nuheimer Str}\ }\textbf {\bibinfo {volume} {19}},\ \bibinfo {pages} {64289}
  (\bibinfo {year} {2008})}\BibitemShut {NoStop}%
\bibitem [{\citenamefont {England}\ \emph {et~al.}(2005)\citenamefont
  {England}, \citenamefont {Rosenzweig}, \citenamefont {Andonian},
  \citenamefont {Musumeci}, \citenamefont {Travish},\ and\ \citenamefont
  {Yoder}}]{england2005sextupole}%
  \BibitemOpen
  \bibfield  {author} {\bibinfo {author} {\bibfnamefont {R.}~\bibnamefont
  {England}}, \bibinfo {author} {\bibfnamefont {J.}~\bibnamefont {Rosenzweig}},
  \bibinfo {author} {\bibfnamefont {G.}~\bibnamefont {Andonian}}, \bibinfo
  {author} {\bibfnamefont {P.}~\bibnamefont {Musumeci}}, \bibinfo {author}
  {\bibfnamefont {G.}~\bibnamefont {Travish}}, \ and\ \bibinfo {author}
  {\bibfnamefont {R.}~\bibnamefont {Yoder}},\ }\href@noop {} {\bibfield
  {journal} {\bibinfo  {journal} {Physical Review Special Topics-Accelerators
  and Beams}\ }\textbf {\bibinfo {volume} {8}},\ \bibinfo {pages} {012801}
  (\bibinfo {year} {2005})}\BibitemShut {NoStop}%
\bibitem [{\citenamefont {Pompili}\ \emph {et~al.}(2016)\citenamefont
  {Pompili}, \citenamefont {Anania}, \citenamefont {Bellaveglia}, \citenamefont
  {Biagioni}, \citenamefont {Castorina}, \citenamefont {Chiadroni},
  \citenamefont {Cianchi}, \citenamefont {Croia}, \citenamefont {Giovenale},
  \citenamefont {Ferrario}, \citenamefont {Filippi}, \citenamefont {Gallo},
  \citenamefont {Gatti}, \citenamefont {Giorgianni}, \citenamefont {Giribono},
  \citenamefont {Li}, \citenamefont {Lupi}, \citenamefont {Mostacci},
  \citenamefont {Petrarca}, \citenamefont {Piersanti}, \citenamefont {Pirro},
  \citenamefont {Romeo}, \citenamefont {Scifo}, \citenamefont {Shpakov},
  \citenamefont {Vaccarezza},\ and\ \citenamefont {Villa}}]{eos_jitter}%
  \BibitemOpen
  \bibfield  {author} {\bibinfo {author} {\bibfnamefont {R.}~\bibnamefont
  {Pompili}}, \bibinfo {author} {\bibfnamefont {M.~P.}\ \bibnamefont {Anania}},
  \bibinfo {author} {\bibfnamefont {M.}~\bibnamefont {Bellaveglia}}, \bibinfo
  {author} {\bibfnamefont {A.}~\bibnamefont {Biagioni}}, \bibinfo {author}
  {\bibfnamefont {G.}~\bibnamefont {Castorina}}, \bibinfo {author}
  {\bibfnamefont {E.}~\bibnamefont {Chiadroni}}, \bibinfo {author}
  {\bibfnamefont {A.}~\bibnamefont {Cianchi}}, \bibinfo {author} {\bibfnamefont
  {M.}~\bibnamefont {Croia}}, \bibinfo {author} {\bibfnamefont {D.~D.}\
  \bibnamefont {Giovenale}}, \bibinfo {author} {\bibfnamefont {M.}~\bibnamefont
  {Ferrario}}, \bibinfo {author} {\bibfnamefont {F.}~\bibnamefont {Filippi}},
  \bibinfo {author} {\bibfnamefont {A.}~\bibnamefont {Gallo}}, \bibinfo
  {author} {\bibfnamefont {G.}~\bibnamefont {Gatti}}, \bibinfo {author}
  {\bibfnamefont {F.}~\bibnamefont {Giorgianni}}, \bibinfo {author}
  {\bibfnamefont {A.}~\bibnamefont {Giribono}}, \bibinfo {author}
  {\bibfnamefont {W.}~\bibnamefont {Li}}, \bibinfo {author} {\bibfnamefont
  {S.}~\bibnamefont {Lupi}}, \bibinfo {author} {\bibfnamefont {A.}~\bibnamefont
  {Mostacci}}, \bibinfo {author} {\bibfnamefont {M.}~\bibnamefont {Petrarca}},
  \bibinfo {author} {\bibfnamefont {L.}~\bibnamefont {Piersanti}}, \bibinfo
  {author} {\bibfnamefont {G.~D.}\ \bibnamefont {Pirro}}, \bibinfo {author}
  {\bibfnamefont {S.}~\bibnamefont {Romeo}}, \bibinfo {author} {\bibfnamefont
  {J.}~\bibnamefont {Scifo}}, \bibinfo {author} {\bibfnamefont
  {V.}~\bibnamefont {Shpakov}}, \bibinfo {author} {\bibfnamefont
  {C.}~\bibnamefont {Vaccarezza}}, \ and\ \bibinfo {author} {\bibfnamefont
  {F.}~\bibnamefont {Villa}},\ }\href
  {http://stacks.iop.org/1367-2630/18/i=8/a=083033} {\bibfield  {journal}
  {\bibinfo  {journal} {New Journal of Physics}\ }\textbf {\bibinfo {volume}
  {18}},\ \bibinfo {pages} {083033} (\bibinfo {year} {2016})}\BibitemShut
  {NoStop}%
\bibitem [{\citenamefont {Faure}\ \emph {et~al.}(2004)\citenamefont {Faure},
  \citenamefont {Glinec}, \citenamefont {Pukhov}, \citenamefont {Kiselev} \emph
  {et~al.}}]{faure2004laser}%
  \BibitemOpen
  \bibfield  {author} {\bibinfo {author} {\bibfnamefont {J.}~\bibnamefont
  {Faure}}, \bibinfo {author} {\bibfnamefont {Y.}~\bibnamefont {Glinec}},
  \bibinfo {author} {\bibfnamefont {A.}~\bibnamefont {Pukhov}}, \bibinfo
  {author} {\bibfnamefont {S.}~\bibnamefont {Kiselev}},  \emph {et~al.},\
  }\href@noop {} {\bibfield  {journal} {\bibinfo  {journal} {Nature}\ }\textbf
  {\bibinfo {volume} {431}},\ \bibinfo {pages} {541} (\bibinfo {year}
  {2004})}\BibitemShut {NoStop}%
\bibitem [{\citenamefont {Geddes}\ \emph {et~al.}(2004)\citenamefont {Geddes},
  \citenamefont {Toth}, \citenamefont {Van~Tilborg}, \citenamefont {Esarey}
  \emph {et~al.}}]{geddes2004high}%
  \BibitemOpen
  \bibfield  {author} {\bibinfo {author} {\bibfnamefont {C.}~\bibnamefont
  {Geddes}}, \bibinfo {author} {\bibfnamefont {C.}~\bibnamefont {Toth}},
  \bibinfo {author} {\bibfnamefont {J.}~\bibnamefont {Van~Tilborg}}, \bibinfo
  {author} {\bibfnamefont {E.}~\bibnamefont {Esarey}},  \emph {et~al.},\
  }\href@noop {} {\bibfield  {journal} {\bibinfo  {journal} {Nature}\ }\textbf
  {\bibinfo {volume} {431}},\ \bibinfo {pages} {538} (\bibinfo {year}
  {2004})}\BibitemShut {NoStop}%
\bibitem [{\citenamefont {Litos}\ \emph {et~al.}(2014)\citenamefont {Litos},
  \citenamefont {Adli}, \citenamefont {An}, \citenamefont {Clarke},
  \citenamefont {Clayton}, \citenamefont {Corde}, \citenamefont {Delahaye},
  \citenamefont {England}, \citenamefont {Fisher}, \citenamefont {Frederico}
  \emph {et~al.}}]{litos2014high}%
  \BibitemOpen
  \bibfield  {author} {\bibinfo {author} {\bibfnamefont {M.}~\bibnamefont
  {Litos}}, \bibinfo {author} {\bibfnamefont {E.}~\bibnamefont {Adli}},
  \bibinfo {author} {\bibfnamefont {W.}~\bibnamefont {An}}, \bibinfo {author}
  {\bibfnamefont {C.}~\bibnamefont {Clarke}}, \bibinfo {author} {\bibfnamefont
  {C.}~\bibnamefont {Clayton}}, \bibinfo {author} {\bibfnamefont
  {S.}~\bibnamefont {Corde}}, \bibinfo {author} {\bibfnamefont
  {J.}~\bibnamefont {Delahaye}}, \bibinfo {author} {\bibfnamefont
  {R.}~\bibnamefont {England}}, \bibinfo {author} {\bibfnamefont
  {A.}~\bibnamefont {Fisher}}, \bibinfo {author} {\bibfnamefont
  {J.}~\bibnamefont {Frederico}},  \emph {et~al.},\ }\href@noop {} {\bibfield
  {journal} {\bibinfo  {journal} {Nature}\ }\textbf {\bibinfo {volume} {515}},\
  \bibinfo {pages} {92} (\bibinfo {year} {2014})}\BibitemShut {NoStop}%
\end{thebibliography}%
\bibliographystyle{apsrev4-1}

\end{document}